\documentclass[12pt,oneside]{amsart}
\usepackage[latin9]{inputenc}
\usepackage{geometry}
\usepackage{mathtools}
\usepackage{amstext}
\usepackage{amsthm}
\usepackage{amssymb}
\usepackage{mathdots}
\usepackage{graphicx}
\usepackage{esint}
\usepackage[unicode=true,pdfusetitle,
 bookmarks=true,bookmarksnumbered=true,bookmarksopen=true,bookmarksopenlevel=1,
 breaklinks=false,pdfborder={0 0 1},backref=section,colorlinks=false]
 {hyperref}
\hypersetup{
 hypertex,bookmarksnumbered=true}

\makeatletter
\numberwithin{equation}{section}
\numberwithin{figure}{section}
\theoremstyle{plain}
\newtheorem{thm}{\protect\theoremname}
\theoremstyle{definition}
\newtheorem{defn}[thm]{\protect\definitionname}
\theoremstyle{definition}
\newtheorem{example}[thm]{\protect\examplename}

\makeatother

\providecommand{\definitionname}{Definition}
\providecommand{\examplename}{Example}
\providecommand{\theoremname}{Theorem}

\begin{document}
\title{Circuit Synthesis based on Prescribed Lagrangian}
\author{Alexander Figotin}
\address{University of California at Irvine, CA 92967}
\email{afigotin@uci.edu}
\begin{abstract}
We advance here an algorithm of a synthesis of an electric circuit
based on prescribed quadratic Lagrangian. That is the circuit evolution
equations are equivalent to the relevant Euler-Lagrange equations.
The proposed synthesis is a systematic approach that allows to realize
any finite dimensional physical system described by a Lagrangian in
a lossless electric circuit so that their evolution equations are
equivalent. The synthesized circuit is composed of (i) capacitors
and inductors of positive or negative values for the respective capacitances
and inductances, and (ii) gyrators. The circuit topological design
is based on the set of $LC$ fundamental loops (f-loops) that are
coupled by $GLC$-links each of which is a serially connected gyrator,
capacitor and inductor. The set of independent variables of the underlying
Lagrangian is identified with f-loop charges defined as the time integrals
of the corresponding currents. The EL equations for all f-loops account
for the Kirchhoff voltage law whereas the Kirchhoff current law holds
naturally as consequence of the setup of the coupled f-loops and the
corresponding charges and currents. The proposed synthesis in particular
provides for efficient implementation of the desired spectral properties
in an electric circuit. The synthesis provides also a way to realize
arbitrary mutual capacitances and inductances through elementary capacitors
and inductors of positive or negative respective capacitances and
inductances.
\end{abstract}

\keywords{Synthesis, electric network, electric circuit, gyrator, Lagrangian,
mutual inductance, mutual capacitance, spectral properties.}
\maketitle

\section{Introduction}

The Lagrangian formalism for electric circuits (networks) is a well-known
subject in electrical engineering. Illustrating examples of constructing
the Lagrangian for rather simple circuits can be found in some monographs
on the variational principles of mechanics, \cite[9]{GantM}, \cite[15]{Wells}.
A growing interest to systematic studies of different aspects of the
Lagrangian formalism for electric networks motivated a number of studies
conducted in the last three decades. To name a few, the general Lagrangian
framework for a broad class of electrical networks, with or without
switches, have been proposed in \cite{Scher}. The Kirchhoff current
law emerges there as a set of constraints for the corresponding Lagrangian
system while the Euler\textendash Lagrange (EL) equations can be interpreted
as the Kirchhoff voltage law. In \cite{Ume16} the Lagrangian method
is used to construct a dual to a non-planar circuit. In \cite{CleSch}
the authors study the relation between the Lagrangian and the Hamiltonian
formalisms for $LC$ circuits. The Hamiltonian formalism of electric
networks with gyrators is considered in \cite{Masc},

The focus of this work is on a certain canonical realization of a
quadratic Lagrangian in a lossless electric network composed of (i)
capacitors and inductors of positive or negative capacitance and inductance,
and (ii) gyrators. We have succeeded in constructing such a realization
and refer to it as \emph{canonical $GLC$-network}. This canonical
$GLC$-network can be viewed topologically as a set of $LC$-loops
coupled to each other by serial combination that involves at least
one of the three basis elements (capacitor, inductor and gyrator)
as illustrated in Fig. \ref{fig:cir-2loop-cp1}. Importantly, capacitances
and inductances of all involved respective capacitors and inductors
can be either positive or negative. Capacitors and inductors of respective
negative capacitance and negative inductance are commonly used in
modern electronics. They after are realized based on operational amplifiers
concisely reviewed in Section \ref{subsec:neg-ICL}.

To illustrate an idea of the canonical $GLC$-network we develop it
first for the simplest case of a quadratic Lagrangian for two variables
defined by
\begin{gather}
\mathcal{L}=\frac{1}{2}\sum_{k,m=1}^{2}\alpha_{km}\dot{Q}_{k}\dot{Q}_{m}+\sum_{k,m=1}^{2}\theta_{km}Q_{m}\dot{Q}_{k}-\frac{1}{2}\sum_{k,m=1}^{2}\eta_{km}Q_{k}Q_{m},\quad\dot{Q}_{k}=\partial_{t}Q_{k},\label{eq:2Loop1a}\\
\alpha_{km}=\alpha_{mk},\quad\theta_{km}=-\theta_{mk},\quad\eta_{km}=\eta{}_{mk},\label{eq:2Loop1b}
\end{gather}
where all coefficients $\alpha_{km}$, $\theta_{km}$ and $\eta_{km}$
are real-valued quantities. The initial step in the construction of
the canonical $GLC$-network is to associate the variables $Q_{1}$
and $Q_{2}$ with charges for respectively two f-loops (see Section
\ref{subsec:net-top}) and then couple them as required. With that
in mind and with understanding that coupling of two f-loops must come
through sharing a branch we use the following elementary identity
$2ab=\left(a+b\right)^{2}-a^{2}-b^{2}$ and recast the Lagrangian
(\ref{eq:2Loop1a}) and (\ref{eq:2Loop1b}) as follows

\begin{equation}
\mathcal{L}=\mathcal{L}_{1}+\mathcal{L}_{2}+\mathcal{L}_{\mathrm{int}},\label{eq:2Loop2a}
\end{equation}
where the Lagrangian components $\mathcal{L}_{1}$, $\mathcal{L}_{1}$
and $\mathcal{L}_{\mathrm{int}}$ are defined by the following expressions

\begin{gather}
\mathcal{L}_{1}=\frac{L_{1}\left(\partial_{t}Q_{1}\right)^{2}}{2}-\frac{\left(Q_{1}\right)^{2}}{2C_{1}},\quad\mathcal{L}_{2}=\frac{L_{2}\left(\partial_{t}Q_{2}\right)^{2}}{2}-\frac{\left(Q_{2}\right)^{2}}{2C_{2}},\label{eq:2Loop2b}
\end{gather}
\begin{gather}
\mathcal{L}_{\mathrm{int}}=G_{12}\left(Q_{1}\partial_{t}Q_{2}-Q_{2}\partial_{t}Q_{1}\right)+\frac{L_{12}\left(\partial_{t}\left(Q_{1}+Q_{2}\right)\right)^{2}}{2}-\frac{\left(Q_{1}+Q_{2}\right)^{2}}{2C_{12}},\label{eq:2Loop2c}
\end{gather}
and parameters $L_{1}$, $L_{2}$, $L_{12}$, $C_{1}$, $C_{2}$,
$C_{12}$ and $G_{12}$ are defined by the following equations 
\begin{equation}
L_{1}=\alpha_{11}-\alpha_{12},\quad L_{2}=\alpha_{22}-\alpha_{21},\quad L_{12}=\alpha_{12},\label{eq:2Loop2d}
\end{equation}
\begin{equation}
C_{1}=\left(\eta_{11}-\eta_{12}\right)^{-1},\quad C_{2}=\left(\eta_{22}-\eta_{21}\right)^{-1},\quad C_{12}=\eta_{12}^{-1},\quad G{}_{12}=\theta_{21}.\label{eq:2Loop2e}
\end{equation}
It is a straightforward exercise to verify the Lagrangian expressions
defined by equations (\ref{eq:2Loop1a}) and (\ref{eq:2Loop1b}) on
one hand and equations (\ref{eq:2Loop2a})-(\ref{eq:2Loop2e}) on
the other hand are exactly equal. Notice that the form of coupling
between two f-loops described by equations (\ref{eq:2Loop2c}) is
perfectly suited to be associated with what we call $GLC$-link, which
is a branch composed of a serially connected gyrator, capacitor and
inductor as shown in Fig. \ref{fig:cir-2loop-cp1}.

In this case the canonical $GLC$-network consists of exactly two
$LC$-loops coupled by a \emph{gyrator link} that involves at least
one of the elements such as gyrator, capacitor and inductor serially
connected as depicted in Fig. \ref{fig:cir-2loop-cp1}. The representation
of the network parameters there in terms of the coefficients $\alpha_{km}$,
$\theta_{km}$ and $\eta_{km}$ of the quadratic Lagrangian defined
by equations (\ref{eq:2Loop1a}) and (\ref{eq:2Loop1b}) is as follows:
\begin{figure*}[h]
\centering{}\includegraphics[scale=0.7]{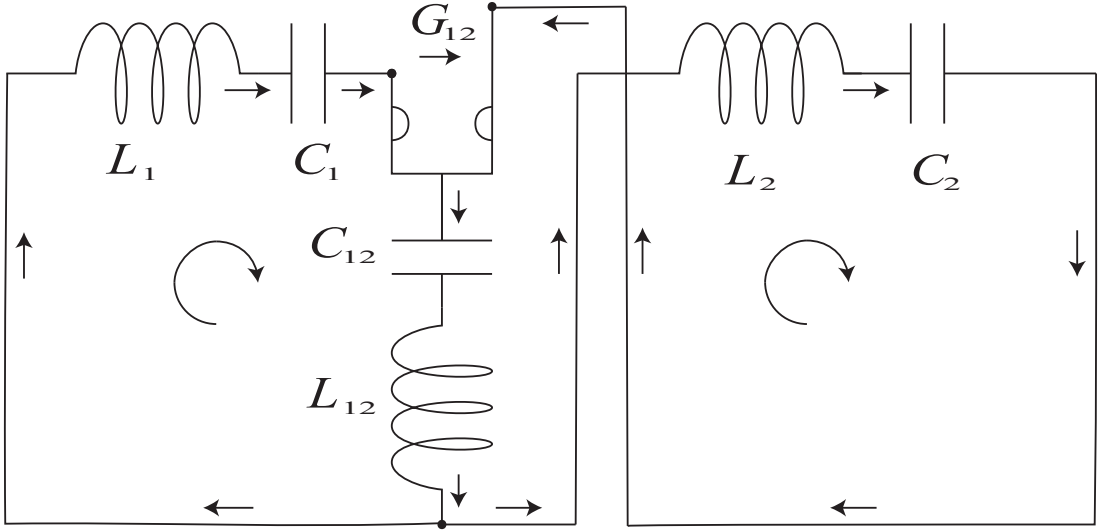}\caption{\label{fig:cir-2loop-cp1} The network of two oriented $LC$-loops
that are coupled by $GLC$-link. The link is composed of ($k$) a
gyrator of the gyrator resistance $G_{12}$, (ii) a capacitor of the
capacitance $C_{12}$ and (iii) an inductor of inductance $L_{12}$
connected in a series. The values of involved elements are defined
by equations (\ref{eq:2Loop2d}) and (\ref{eq:2Loop2e}). Notice the
difference between the left and the right connections for the gyrator
and $LC$-loops. It is explained by the non-reciprocity of the gyrator
and is designed to be consistent with (i) the standard port assignment
and selection of positive directions for the loop currents and the
gyrator; (ii) the sign of gyration resistance as shown in Fig. \ref{fig:cir-gyr}
and equations (\ref{eq:crvc1b}).}
\end{figure*}

A simple procedure for the construction of the canonical $GLC$-loop
Lagrangian form of the original general quadratic Lagrangian is provided
in Section \ref{sec:prep-syn}. This canonical $GLC$-loop Lagrangian
leads straightforwardly to the synthesis of the canonical $GLC$-network
associated with the Lagrangian.

\section{Preparation to the circuit synthesis\label{sec:prep-syn}}

We are interested in lossless circuits, that is there no resistors
are involved. We name such circuits \emph{$GLC$-circuits} and define
them as follows.
\begin{defn}[$GLC$-circuit]
\label{def:GLC-cir} $GLC$-circuit is defined as any circuit composed
of (i) capacitors and inductors of positive or negative values for
the respective capacitances and inductances; (ii) gyrators.
\end{defn}

Let the Lagrangian be of general quadratic form
\begin{gather}
\mathcal{L}=\frac{1}{2}\dot{Q}^{\mathrm{T}}\alpha\dot{Q}+\dot{Q}^{\mathrm{T}}\theta Q-\frac{1}{2}Q^{\mathrm{T}}\eta Q,\quad\dot{Q}=\partial_{t}Q,\quad Q=\left[Q_{1},Q_{2},\ldots,Q_{N}\right]{}^{\mathrm{T}},\label{eq:LagQ1a}
\end{gather}
or
\begin{gather}
\mathcal{L}=\frac{1}{2}\sum_{k,m=1}^{N}\alpha_{km}\dot{Q}_{k}\dot{Q}_{m}+\sum_{k,m=1}^{N}\theta_{km}Q_{m}\dot{Q}_{k}-\frac{1}{2}\sum_{k,m=1}^{N}\eta_{km}Q_{k}Q_{m},\label{eq:LagQ1b}
\end{gather}
where $\alpha=\left\{ \alpha_{km}\right\} $, $\theta=\left\{ \theta_{km}\right\} $
and $\eta=\left\{ \eta_{km}\right\} $ are $N\times N$ matrices satisfying
\begin{gather}
\alpha_{km}=\alpha_{mk},\quad\theta_{km}=-\theta_{mk},\quad\eta_{km}=\eta{}_{mk},\quad1\leq k,m\leq N.\label{eq:LagQ1c}
\end{gather}
\begin{figure*}[h]
\centering{}\includegraphics[scale=0.6]{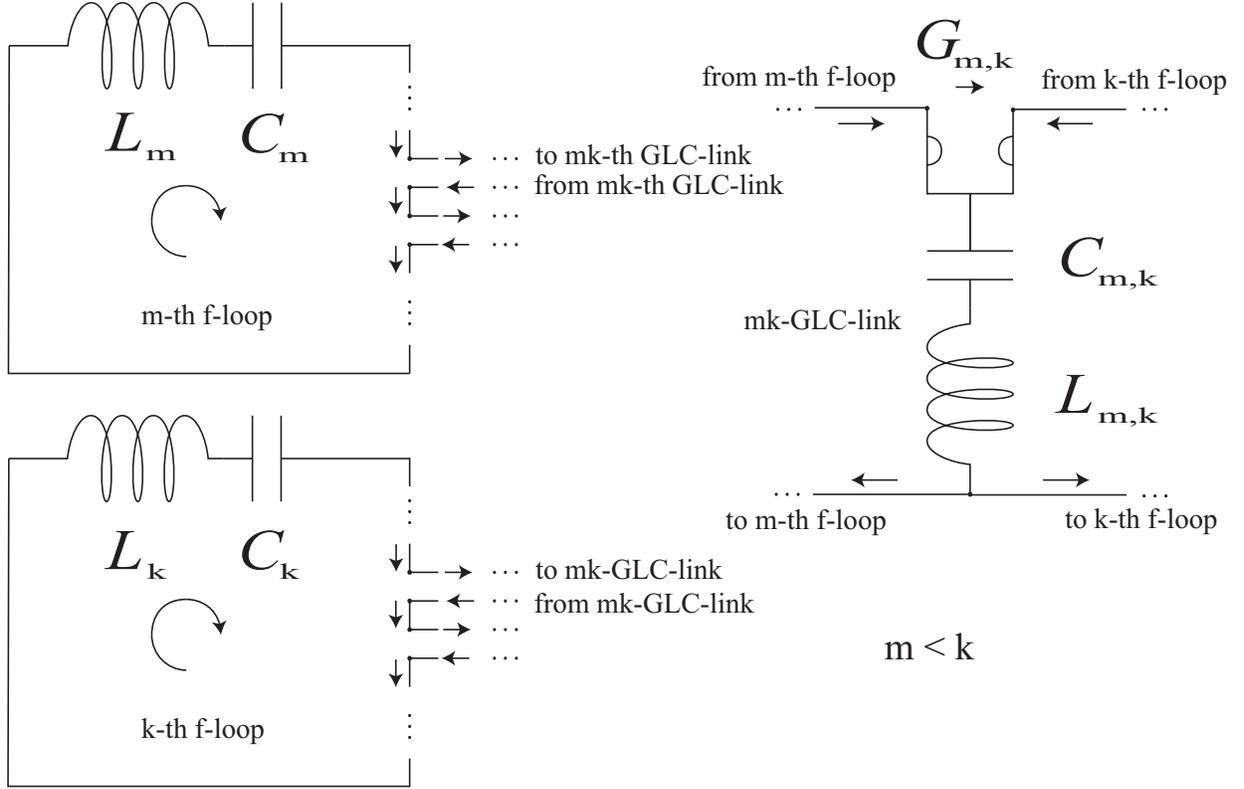}\caption{\label{fig:loop-gyr} This figure provides graphical support to the
circuit synthesis by showing how the $m$-th and the $k$-th f-loops
are coupled by the $GLC$-link of the index $m,k$. It is understood
that any other pair of f-loops can be coupled if needed similarly.
Notice that under assumption $m<k$ the $m$-th and $k$-th f-loop
are connected respectively to the left and to right shoulders of $GLC$-link.
The arrows indicate the positive directions of the currents. All shown
connections are designed to be consistent with (i) the standard port
assignment and selection of positive directions for the loop currents
and the gyrator; (ii) the sign of gyration resistance as shown in
Fig. \ref{fig:cir-gyr} and equations (\ref{eq:crvc1b}). }
\end{figure*}

The first step of the circuit synthesis based on the prescribed Lagrangian
defined by equations (\ref{eq:LagQ1a}) - (\ref{eq:LagQ1c}) is to
transform it into a particular form suited for its circuit implementation.
This transformation starts with the interpretation of the coordinates
$Q_{m}$ and the corresponding generalized velocities $v_{m}=\partial_{t}Q_{m}$,
$1\leq m\leq N$ as respectively the $m$-th f-loop charges and currents
of the circuit to be constructed. With this interpretation in mind
the terms $\frac{1}{2}\alpha_{mm}\dot{Q}_{m}^{2}$ and $\frac{1}{2}\eta_{mm}Q_{m}^{2}$
can be naturally attributed respectively to the energies of an inductor
of the inductance $L_{m}=\alpha_{mm}$ and a capacitor of the inverse
capacitance $B_{m}=C_{m}^{-1}=\eta_{mm}$ associated with $m$-th
f-loop. The terms $\frac{1}{2}\alpha_{km}\dot{Q}_{k}\dot{Q}_{m}$
and $\frac{1}{2}\eta_{km}Q_{k}Q_{m}$ for $k\neq m$ are commonly
attributed respectively to the energies of mutual inductance $L_{km}=\alpha_{km}$
and mutual capacitance of the inverse capacitance $B_{km}=C_{km}^{-1}=\eta_{mm}$.
Though these terms evidently carry information on mutual coupling
between different f-loops they are not linked directly to branches
of a circuit. To address the problem we use elementary identity $2ab=\left(a+b\right)^{2}-a^{2}-b^{2}$
for $a=Q_{k},\dot{Q}_{k}$ and $b=Q_{m},\dot{Q}_{m}$ and rearrange
the Lagrangian terms as follows
\begin{gather}
\frac{1}{2}\sum_{k,m=1}^{N}\alpha_{km}\dot{Q}_{k}\dot{Q}_{m}=\frac{1}{2}\sum_{k=1}^{N}\left(2\alpha_{kk}-\sum_{m=1}^{N}\alpha_{km}\right)\dot{Q}_{k}^{2}+\frac{1}{2}\sum_{1\leq m<k\leq N}\alpha_{km}\left(\dot{Q}_{k}+\dot{Q}_{m}\right)^{2}.\label{eq:Labcqv1d}
\end{gather}

\begin{gather}
\frac{1}{2}\sum_{k,m=1}^{N}\eta_{km}Q_{k}Q_{m}=\frac{1}{2}\sum_{k=1}^{N}\left(2\eta_{kk}-\sum_{m=1}^{N}\eta_{km}\right)Q_{k}^{2}+\frac{1}{2}\sum_{1\leq m<k\leq N}\left(Q_{k}+Q_{m}\right)^{2},\label{eq:Labcqv1c}
\end{gather}
Notice also
\begin{equation}
\sum_{k,m}\theta_{km}Q_{m}\dot{Q}_{k}=\sum_{1\leq m<k\leq N}\theta_{km}\left(Q_{m}\dot{Q}_{k}-Q_{k}\dot{Q}_{m}\right).\label{eq:Labcqv1e}
\end{equation}
Based on identities (\ref{eq:Labcqv1d})-(\ref{eq:Labcqv1e}) we recast
the original Lagrangian defined by equations (\ref{eq:LagQ1a})-(\ref{eq:LagQ1c})
as

\begin{equation}
\mathcal{L}=\sum_{k=1}^{N}\mathcal{L}_{k}+\sum_{1\leq m<k\leq N}\mathcal{L}_{km},\label{eq:LagGLC1a}
\end{equation}
\begin{equation}
\mathcal{L}_{k}=L_{k}\dot{Q}_{k}^{2}+B_{k}Q_{k}^{2},\;\dot{Q}_{k}=\partial_{t}\left(Q\right),\;1\leq k\leq N,\label{eq:LagGLC1b}
\end{equation}
\begin{gather}
\mathcal{L}_{km}=L_{km}\left(\dot{Q}_{k}+\dot{Q}_{m}\right)^{2}+B_{km}\left(Q_{k}+Q_{m}\right)^{2}+G_{km}\left(Q_{m}\dot{Q}_{k}-Q_{k}\dot{Q}_{m}\right),\;1\leq m<k\leq N,\label{eq:LagGLC1c}
\end{gather}
where inductances $L_{k}$ and $L_{km}$, inverse capacitances $B_{k}$
and $B_{km}$, and gyration resistances $G_{km}$ are defined by
\begin{gather}
L_{k}=2\alpha_{kk}-\sum_{m=1}^{N}\alpha_{km}=\alpha_{kk}-\sum_{m\neq k}\alpha_{km},\quad1\leq k\leq N;\label{eq:LagGLC1d}
\end{gather}
\begin{equation}
L_{km}=\alpha_{km},\quad1\leq m<k\leq N;\label{eq:LagGLC1da}
\end{equation}
\begin{gather}
B_{k}=C_{k}^{-1}=2\eta_{kk}-\sum_{m=1}^{N}\eta_{km}=\eta_{kk}-\sum_{m\neq k}\eta_{km},\quad1\leq k\leq N;\label{eq:LagGLC1e}
\end{gather}
\begin{equation}
B_{km}=C_{km}^{-1}=\eta{}_{km},\quad G_{km}=\theta_{km},\quad1\leq m<k\leq N,\label{eq:LagGLC1ea}
\end{equation}
We refer to the Lagrangian representation (\ref{eq:LagGLC1a})-(\ref{eq:LagGLC1c})
as the \emph{loop Lagrangian}. The point of the above transformation
leading to the loop Lagrangian is to replace products $\dot{Q}_{k}\dot{Q}_{m}$
and $Q_{k}Q_{m}$ with respectively $\left(\dot{Q}_{k}+\dot{Q}_{m}\right)^{2}$
and $\left(Q_{k}+Q_{m}\right)^{2}$ accompanied by the proper modification
of the coefficients before $\dot{Q}_{k}^{2}$ and $Q_{m}^{2}$. With
that done we notice first that the Lagrangian $\mathcal{L}_{k}$ defined
by equations (\ref{eq:LagGLC1b}) can naturally be associated with
$k$-th f-loop of the circuit to be constructed, see Section \ref{subsec:cir-elem}
and equations (\ref{eq:cirvc3a}), (\ref{eq:cirvc3b}) there. As to
the Lagrangian $\mathcal{L}_{km}$ defined by equations (\ref{eq:LagGLC1c})
it is perfectly suited to be attributed to what we call a \emph{$GLC$-link}
which is a connector between $k$-th and $m$-th f-loops defined as
follows.
\begin{defn}[$GLC$-link]
\label{def:GLC-link} $GLC$-link, shown on the right side of Fig.
\ref{fig:loop-gyr}, is a branch composed of a circuit made of (i)
a gyrator of the gyrator resistance $G$, (ii) a capacitor of the
capacitance $C$ and (iii) an inductor of inductance $L$ connected
in a series. Some of the quantities $G$, $B=C^{-1}$and $L$ are
allowed to be zero but at least one them is required to be non-zero.
In other words, $GLC$-link can be composed of 3, 2 or 1 elements.
For instance, it can just a gyrator or capacitor, or it can be a combination
of a capacitor and inductance.
\end{defn}

Indeed, equations (\ref{eq:LagGLC1c}) suggest that the charge and
the corresponding current associated with the $GLC$-link are respectively
$Q_{k}+Q_{m}$ and $\partial_{t}\left(Q_{k}+Q_{m}\right)$ and the
gyroscopic interaction between $k$-th and $m$-th f-loops is accounted
for by the term involving gyration resistances $G_{km}$.

Notice that it might happen that for particular values of indexes
$\mathcal{L}_{k}=0$ and/or $\mathcal{L}_{km}=0$, implying effectively
that the corresponding circuit elements have to be eliminated. In
particular, the possibility of $\mathcal{L}_{km}=0$ can result in
a circuit which is a composition of two or more totally disconnected
components. In view of these factors we want to limit naturally our
considerations to Lagrangians that yield connected circuits only.
With that in mind we introduce the concepts of \emph{connected f-loops}
and of \emph{admissible Lagrangian} as follows.
\begin{defn}[connected f-loops]
\label{def:confloop} Using the expressions (\ref{eq:LagGLC1c})
for the Lagrangian component $\mathcal{L}_{km}$ valid for $1\leq m<k\leq N$
we define $\mathcal{L}_{km}$ for every pair $k\neq m$ by the following
equality
\begin{equation}
\mathcal{L}_{km}=\mathcal{L}_{sp},\;s=\min\left\{ k,m\right\} ,\;p=\max\left\{ k,m\right\} .\label{eq:admLag1b}
\end{equation}
We define then for $k\neq m$ the $k$-th and the $m$-th f-loops
to be \emph{directly connected} if $\mathcal{L}_{km}\neq0$. For $k\neq m$
the $k$-th and the $m$-th f-loops are called \emph{connected} if
the following is true. There exists a set $S_{km}=\left\{ s_{1},\ldots,s_{q}\right\} $
of distinct indexes $s_{i}\neq k,m$ such that for every pair $\left\{ s_{i},s_{i+1}\right\} $
the corresponding f-loops are directly connected for $0\leq i\leq q$
assuming that $s_{0}=k$ and $s_{q+1}=m$. In other words, the $k$-th
and the $m$-th f-loops are connected if there exists a ``path''
from $k$ to $m$ in the set of indexes such the end points of each
of its segments correspond to directly connected f-loops.
\end{defn}

Having defined connected f-loops we proceed with the following definition.
\begin{defn}[admissible Lagrangian]
\label{def:admLag} We refer to the loop Lagrangian defined by equations
(\ref{eq:LagGLC1a})-(\ref{eq:LagGLC1c}) as admissible if (i) for
every pair of different indexes $\left\{ k,m\right\} $ corresponding
f-loops are connected; (ii) every $k$-th f-loop Lagrangian $\mathcal{L}_{k}$
defined by equations (\ref{eq:LagGLC1a}) and (\ref{eq:LagGLC1b})
is not zero, that is
\begin{equation}
\mathcal{L}_{k}\neq0,\;1\leq k\leq N.\label{eq:admLag1a}
\end{equation}
\end{defn}

\section{The circuit synthesis\label{sec:cir-syn}}

The circuit synthesis described in this sections is based on the results
of Section \ref{sec:prep-syn}. It assumes that: (i) the Lagrangian
is admissible according to Definition \ref{def:admLag}, and (ii)
the synthesized circuit is an $GLC$-circuit as in Definition \ref{def:GLC-cir}.
In fact, the synthesis process is rather straightforward and guided
by Fig. \ref{fig:loop-gyr} providing visual support. We still add
to that formal steps of the synthesis algorithm to make sure that
the synthesized $GLC$-circuit implements the desired goals.

The circuit synthesis algorithm can be viewed a process of consequent
modifications of the set of initially isolated f-loops as they get
connected by proper $GLC$-links. The process utilizes the expressions
of the loop Lagrangian components $\mathcal{L}_{k}$ and $\mathcal{L}_{km}$
defined by equations (\ref{eq:LagGLC1a})-(\ref{eq:LagGLC1c}) and
its steps are as follows.

\textbf{The circuit synthesis algorithm}.
\begin{enumerate}
\item Set up the initial state of the f-loops. Based on the Lagrangians
$\mathcal{L}_{k}$, $1\leq k\leq N$ we create the corresponding $LC$-circuits
(f-loops) $\mathsf{F}_{k}$ with the values of inductances $L_{k}$
and capacitances $C_{k}$ defined by equations (\ref{eq:LagGLC1d})
and (\ref{eq:LagGLC1e}). These initially disconnected f-loops $\mathsf{F}_{k}$
constitute the initial state of circuit before the their coupling
process is fully implemented.
\item Create the $GLC$-links to be used for coupling of the f-loops. Based
on non-zero Lagrangians $\mathcal{L}_{km}$, $1\leq m<k\leq N$ we
create $GLC$-links $\mathsf{G}_{km}$ with the values of inductances
$L_{km}$, capacitances $C_{km}$ and gyration resistances $G_{km}$
defined by equations (\ref{eq:LagGLC1da}) and (\ref{eq:LagGLC1ea}).
\item Modify recurrently f-loops $\mathsf{F}_{k}$ by coupling them with
$GLC$-links $\mathsf{G}_{km}$. The modification process starts with
the initial state of the f-loops $\mathsf{F}_{k}$. Its goal is to
add up to the initial state of the $GLC$-circuit one-by-one all non-zero
$GLC$-links $\mathsf{G}_{km}$. Suppose that $\mathsf{F}_{m}^{\prime}$
and $\mathsf{F}_{k}^{\prime}$ for $m<k$ constitute an intermediate
state of the $m$-th and $k$-th loops in the process of adding non-zero
$GLC$-links. We modify them by adding up $GLC$-link $\mathsf{G}_{km}$
as follows. We open up the f-loop $\mathsf{F}_{m}^{\prime}$ by removing
its wire segment and connect the ends of the so open f-loop to the
left shoulder of $GLC$-link $\mathsf{G}_{km}$ as indicated in Fig.
\ref{fig:loop-gyr}. Then we open up the f-loop $\mathsf{F}_{k}^{\prime}$
by removing its wire segment and connect the ends of the so open f-loop
to the right shoulder of $GLC$-link $\mathsf{G}_{km}$ as indicated
in Fig. \ref{fig:loop-gyr}. As the result of the modification we
get new states $\mathsf{F}_{m}^{\prime\prime}$ and $\mathsf{F}_{k}^{\prime\prime}$
for the corresponding f-loops with integrated into them $GLC$-link
$\mathsf{G}_{km}$. Executing the described process until all $GLC$-links
$\mathsf{G}_{km}$ are integrated into the $GLC$-circuit we arrive
at the desired synthesized circuit.
\end{enumerate}
It follows from the circuit synthesis algorithm that the synthesized
$GLC$-circuit is uniquely defined by the prescribed Lagrangian. This
circuit can be viewed as canonical and we name it \emph{canonical
$GLC$-circuit}. Since any $GLC$-circuit can be associated with the
Lagrangian based on equations (\ref{eq:cirvc3a}) and (\ref{eq:cirvc3a})
for its elementary terms the following statements hold.
\begin{thm}[circuit implementation of a Lagrangian system]
\label{thm:Lag-cir} Any finite-dimensional physical system described
by a Lagrangian can be implemented as the canonical $GLC$-circuit
associated with it.
\end{thm}

\begin{thm}[equivalent canonical $GLC$-circuit]
 \label{thm:equi-cancir} Any $GLC$-circuit has an equivalent representation
as the canonical $GLC$-circuit in the sense that the evolution equations
for the both circuits are equivalent.
\end{thm}

A $GLC$-circuit can be different from its canonical $GLC$-circuit.
This is the case when the circuit has a twig which is common to more
than two loops as demonstrated below by an example of simple $GLC$-circuit
that involves only inductors and capacitors.
\begin{example}[canonical $GLC$-circuit]
\label{exa:canGLC} Let us consider a circuit composed of 3 f-loops
with inductances and inverse capacitances respectively $L_{k},B_{k}$,$\:k=1,2,3$
and having a single common branch with inductances and inverse capacitances
respectively $L_{b},B_{b}$. The corresponding circuit Lagrangian
then is
\begin{gather}
\mathcal{L}=\frac{1}{2}\sum_{k=1}^{3}\left(L_{k}\dot{Q}_{k}^{2}-B_{k}Q_{k}^{2}\right)+\frac{L_{\mathrm{b}}}{2}\left(\dot{Q}_{1}+\dot{Q}_{2}+\dot{Q}_{3}\right)^{2}-\frac{B_{\mathrm{b}}}{2}\left(Q_{1}+Q_{2}+Q_{3}\right)^{2}.\label{eq:ex1Lag1a}
\end{gather}
Then according to equations (\ref{eq:LagGLC1d}), (\ref{eq:LagGLC1e}),
(\ref{eq:LagGLC1da}) and (\ref{eq:LagGLC1ea}) the canonical $GLC$-circuit
involves only inductors and capacitors of the following respective
values for inductances and inverse capacitances
\begin{equation}
\mathring{L}_{k}=L_{k}-L_{\mathrm{b}},\quad\mathring{B}_{k}=B_{k}-B_{\mathrm{b}},\quad k=1,2,3,\label{eq:ex1Lag1b}
\end{equation}
\begin{equation}
\mathring{L}_{km}=L_{\mathrm{b}},\quad\mathring{B}_{mk}=B_{\mathrm{b}},\quad1\leq m<k\leq3.\label{eq:ex1Lag1c}
\end{equation}
Notice that if the values of all inductances and capacitances of the
original circuit are positive, that is $L_{k},L_{\mathrm{b}}>0,B_{k},B_{\mathrm{b}}>0$,$\:k=1,2,3$
then according to equations (\ref{eq:ex1Lag1b}) and (\ref{eq:ex1Lag1c})
some of the corresponding values $\mathring{L}_{k}$ and $\mathring{B}_{k}$
for the canonical $GLC$-circuit can evidently be negative.
\end{example}

\section{A Sketch of the Basics of Electric Networks\label{sec:e-net}}

For the sake of self-consistency, we provide in this section basic
information on the basics of the electric network theory and relevant
notations.

Electrical networks is a well established subject represented in many
monographs. We present here basic elements of the electrical network
theory following mostly to \cite[2]{BalBic}, \cite{Cau}, \cite{SesRee}.
The electrical network theory constructions are based on the graph
theory concepts of branches (edges), nodes (vertices) and their incidences.
This approach is efficient in loop (fundamental circuit) analysis
and the determination of independent variables for the Kirchhoff current
and voltage laws - the subjects relevant to our studies here.

We are particularly interested in conservative electrical network
which is a particular case of an electrical network composed of electric
elements of three types: capacitors, inductors and gyrators. We remind
that a capacitor or an inductor are the so-called two-terminal electric
elements whereas a gyrator is four-terminal electric element as discussed
below. We assume that capacitors and inductors can have positive or
negative respective capacitances and inductances.

\subsection{Circuit elements and their voltage-current relationships\label{subsec:cir-elem}}

\begin{figure}[h]
\centering{}\includegraphics[scale=0.35]{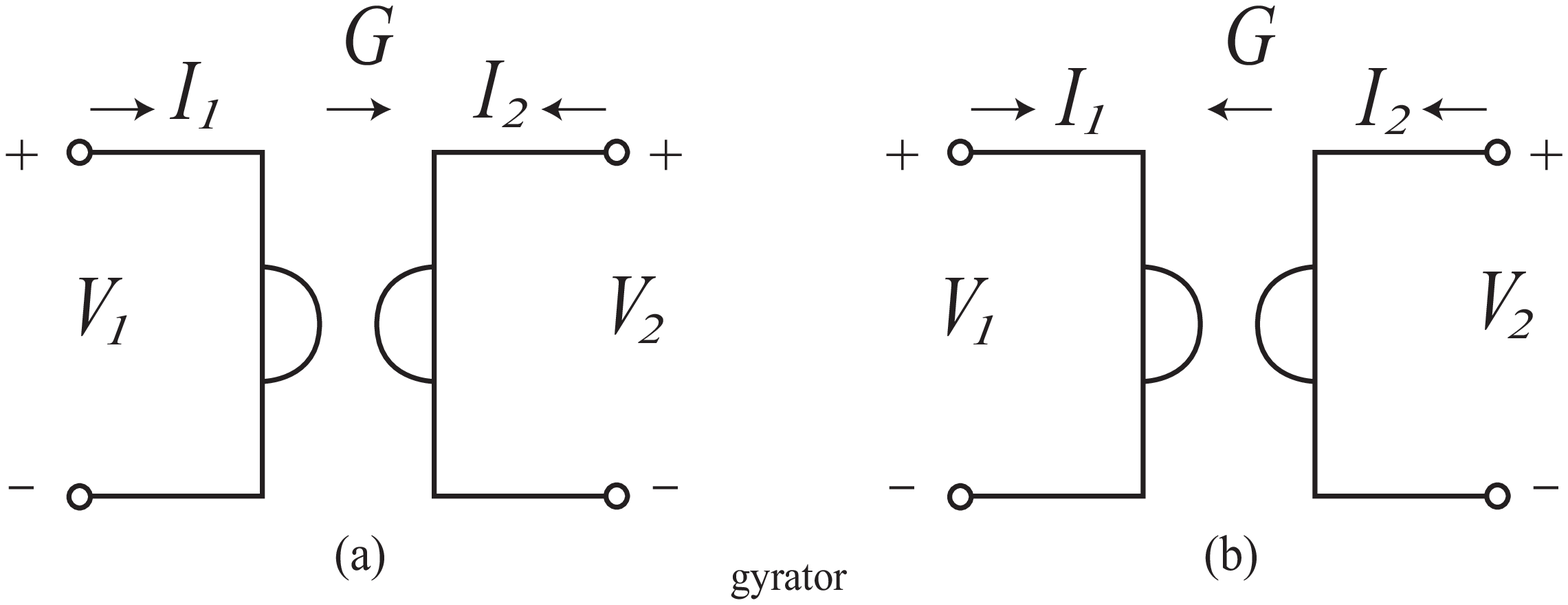}\caption{\label{fig:cir-gyr} Gyrator.}
\end{figure}

The elementary electric network (circuit) elements of interest here
are a \emph{capacitor}, an \emph{inductor}, a \emph{resistor} and
a \emph{gyrator}, \cite[1.5, 2.6]{BalBic}, \cite[App.5.4]{Cau},
\cite[10]{Iza}. These elements are characterized by the relevant
\emph{voltage-current relationships}. These relationships for the
capacitor, inductor and resistor are respectively as follows \cite[1.5]{BalBic},
\cite[3-Circuit theory]{Rich}, \cite[1.3]{SesBab}:
\begin{equation}
I=C\partial_{t}V,\quad V=L\partial_{t}I,\quad V=RI,\label{eq:cirvc1a}
\end{equation}
where $I$ and $V$ are respectively the \emph{current} and the \emph{voltage},
and real $C$, $L$ and $R$ are called respectively the \emph{capacitance},
the \emph{inductance} and the \emph{resistance}. The voltage-current
relationship for the gyrator depicted in Fig. \ref{fig:cir-gyr} are
\begin{gather}
(a):\:\begin{bmatrix}V_{1}\\
V_{2}
\end{bmatrix}=\begin{bmatrix}-GI_{2}\\
GI_{1}
\end{bmatrix},\quad(b):\:\begin{bmatrix}V_{1}\\
V_{2}
\end{bmatrix}=\begin{bmatrix}GI_{2}\\
-GI_{1}
\end{bmatrix},\label{eq:crvc1b}\\
\nonumber 
\end{gather}
where $I_{1},\:I_{2}$ and $V_{1},\:V_{2}$ are respectively the \emph{currents}
and the \emph{voltages}, and quantity $G$ is called the \emph{gyration
resistance.}

\begin{figure}[h]
\centering{}\includegraphics[scale=0.25]{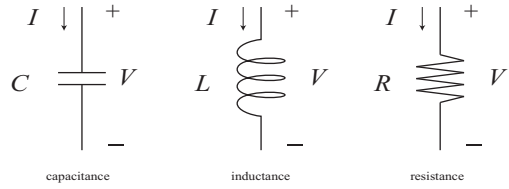}\caption{\label{fig:cir-CLR} Capacitance, inductance and resistance.}
\end{figure}

The common graphic representations of the network elements are depicted
in Figs. \ref{fig:cir-CLR} and \ref{fig:cir-gyr}. The arrow next
to the symbol $G$ in Fig. \ref{fig:cir-gyr} shows the direction
of gyration.

\begin{figure}[h]
\centering{}\includegraphics[scale=0.35]{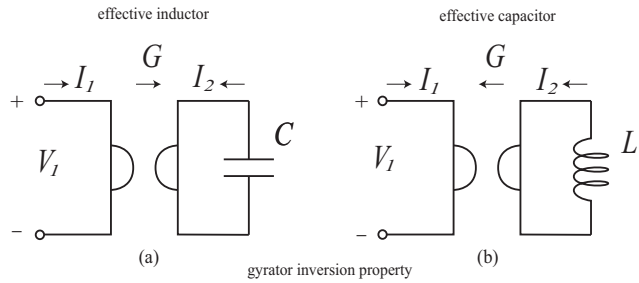}\caption{\label{fig:cir-gyr-inv} (a) Effective inductor; (b) effective capacitor.}
\end{figure}

The gyrator has the so-called inverting property as shown in Fig.
\ref{fig:cir-gyr-inv}, \cite[1.5]{BalBic}, \cite[29.1]{Dorf}, \cite[10]{Iza}.
Namely, when a capacitor or an inductor connected to the output port
of the gyrator it behaves as an inductor or capacitor respectively
with the following effective values
\begin{equation}
L_{\mathrm{ef}}=G^{2}C,\quad C_{\mathrm{ef}}=\frac{L}{G^{2}}.\label{eq:cirvc1d}
\end{equation}

Notice that the voltage-current relationships in equations (\ref{eq:crvc1b})
(b) can be obtained from the same in equations (\ref{eq:crvc1b})
(a) by substituting $-G$ for $G$. The gyrator is a device that accounts
for physical situations in which the reciprocity condition does not
hold. The voltage-current relationships in equations (\ref{eq:crvc1b})
show that the gyrator is a non-reciprocal circuit element. In fact,
it is antireciprocal. Notice, that the gyrator, like the ideal transformer,
is characterized by a single parameter $G$, which is the gyration
resistance. The arrows next to the symbol $G$ in Fig. \ref{fig:cir-gyr}(a)
and (b) show the direction of gyration.

Along with the voltage $V$ and the current $I$ variables we introduce
the \emph{charge }variable $Q$ and the \emph{momentum} (per unit
of charge) variable $P$ by the following formulas
\begin{gather}
Q\left(t\right)=\intop I\left(t\right)\,dt,\quad I\left(t\right)=\partial_{t}Q,\label{eq:cirvc2a}\\
P\left(t\right)=\intop V\left(t\right)\,dt,\quad V\left(t\right)=\partial_{t}P.\label{eq:cirvc2b}
\end{gather}
We introduce also the energy stored variable $W$, \cite[Circuit Theory]{Rich}.
Then the voltage-current relations (\ref{eq:cirvc1a}) and the stored
energy $W$ can be represented as follows:

\begin{gather}
\text{capacitor: }V=\frac{Q}{C},\quad I=\partial_{t}Q=C\partial_{t}V,\label{eq:cirvc2c}\\
Q=CV=C\partial_{t}P;\nonumber \\
W=\frac{1}{2}VQ=\frac{Q^{2}}{2C}=\frac{CV^{2}}{2}=\frac{C\left(\partial_{t}P\right)^{2}}{2};\label{eq:cirvc2d}
\end{gather}

\begin{gather}
\text{inductor: }V=L\partial_{t}I,\quad P=LI=L\partial_{t}Q,\label{eq:cirvc2e}\\
\partial_{t}Q=\frac{P}{L};\nonumber \\
W=\frac{PI}{2}=\frac{LI^{2}}{2}=\frac{L\left(\partial_{t}Q\right)^{2}}{2}=\frac{P^{2}}{2L};\label{eq:cirvc2f}
\end{gather}
\begin{equation}
\text{resistor: }V=RI,\quad P=RQ.\label{eq:cirvc2g}
\end{equation}

The Lagrangian associated with the network elements are as follows
\cite[9]{GantM}, \cite[3]{Rich}:
\begin{equation}
\text{capacitor: }\mathcal{L}=\frac{Q^{2}}{2C},\:\text{inductor: }\mathcal{L}=\frac{L\left(\partial_{t}Q\right)^{2}}{2},\label{eq:cirvc3a}
\end{equation}
\begin{gather}
\text{gyrator: }\mathcal{L}=GQ_{1}\partial_{t}Q_{2},\label{eq:cirvc3b}\\
\mathcal{L}=\frac{G\left(Q_{1}\partial_{t}Q_{2}-Q{}_{2}\partial_{t}Q_{1}\right)}{2}.\nonumber 
\end{gather}
Notice that the difference between two alternatives for the Lagrangian
in equations (\ref{eq:cirvc3b}) is $\frac{1}{2}G\partial_{t}\left(Q_{1}Q_{2}\right)$
which is evidently the complete time derivative. Consequently, the
EL equation are the same for both Lagrangians, see Section \ref{sec:Lag}.

\subsection{Circuits of negative impedance, capacitance and inductance\label{subsec:neg-ICL}}

\begin{figure}[h]
\begin{centering}
\includegraphics[scale=0.45]{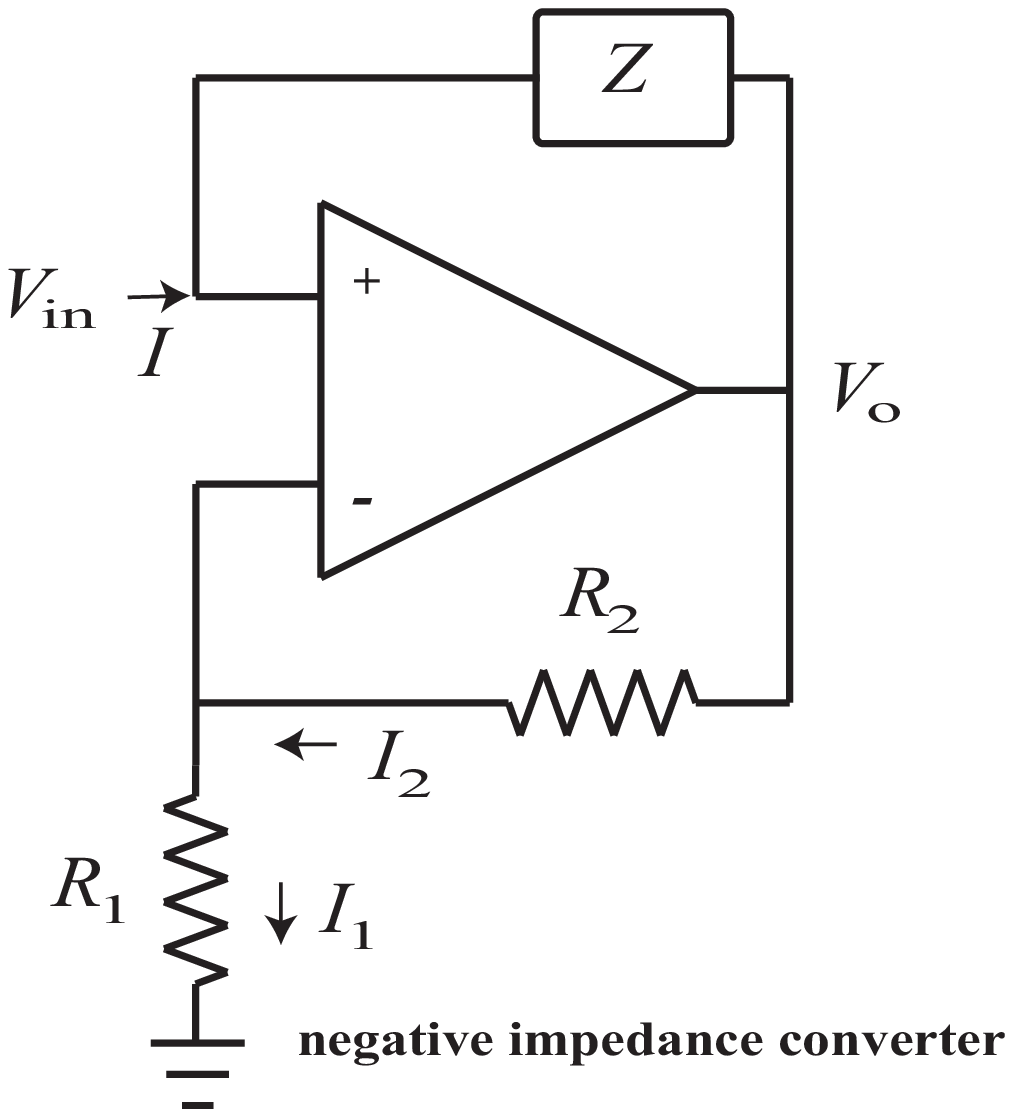}\hspace{1.5cm}\includegraphics[scale=0.45]{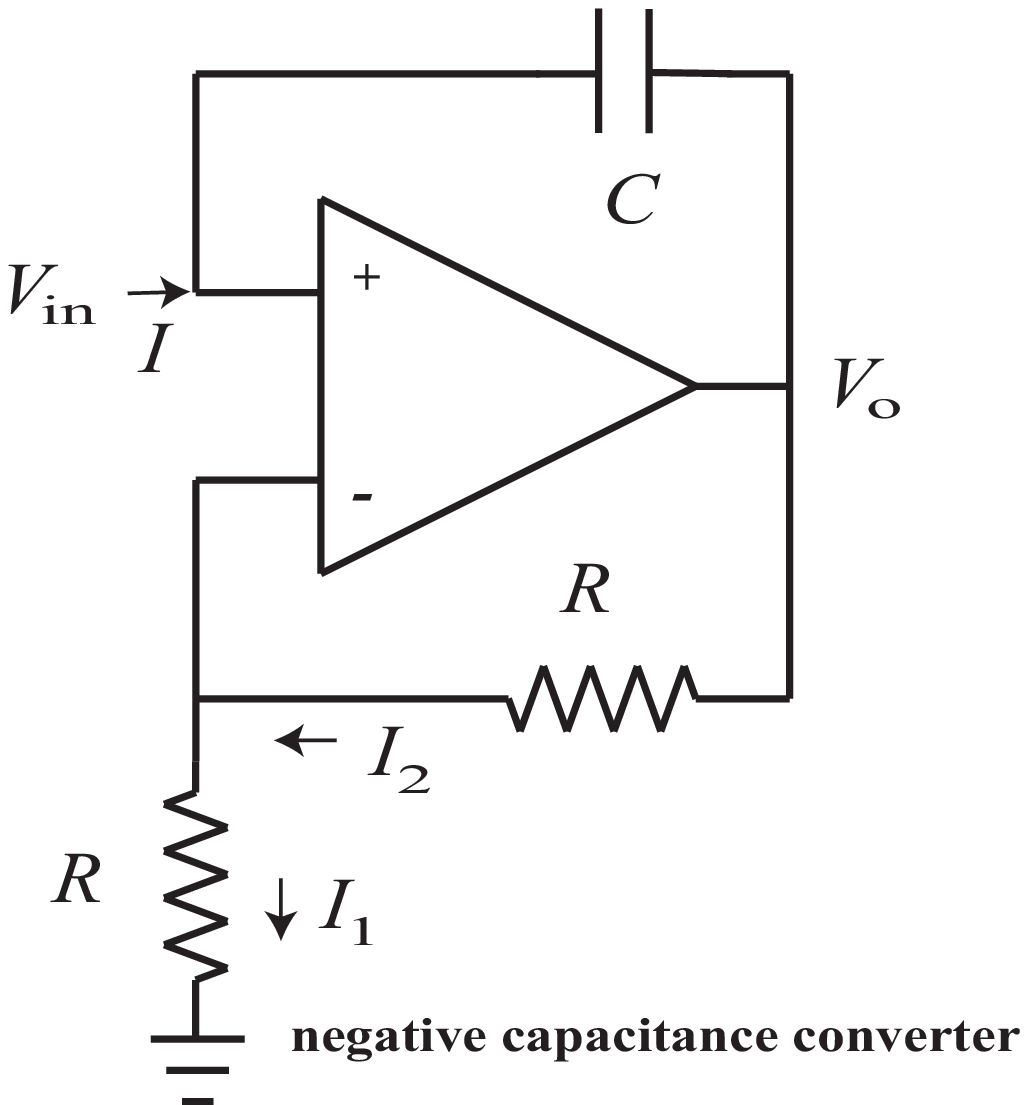}\hspace{1.5cm}\includegraphics[scale=0.45]{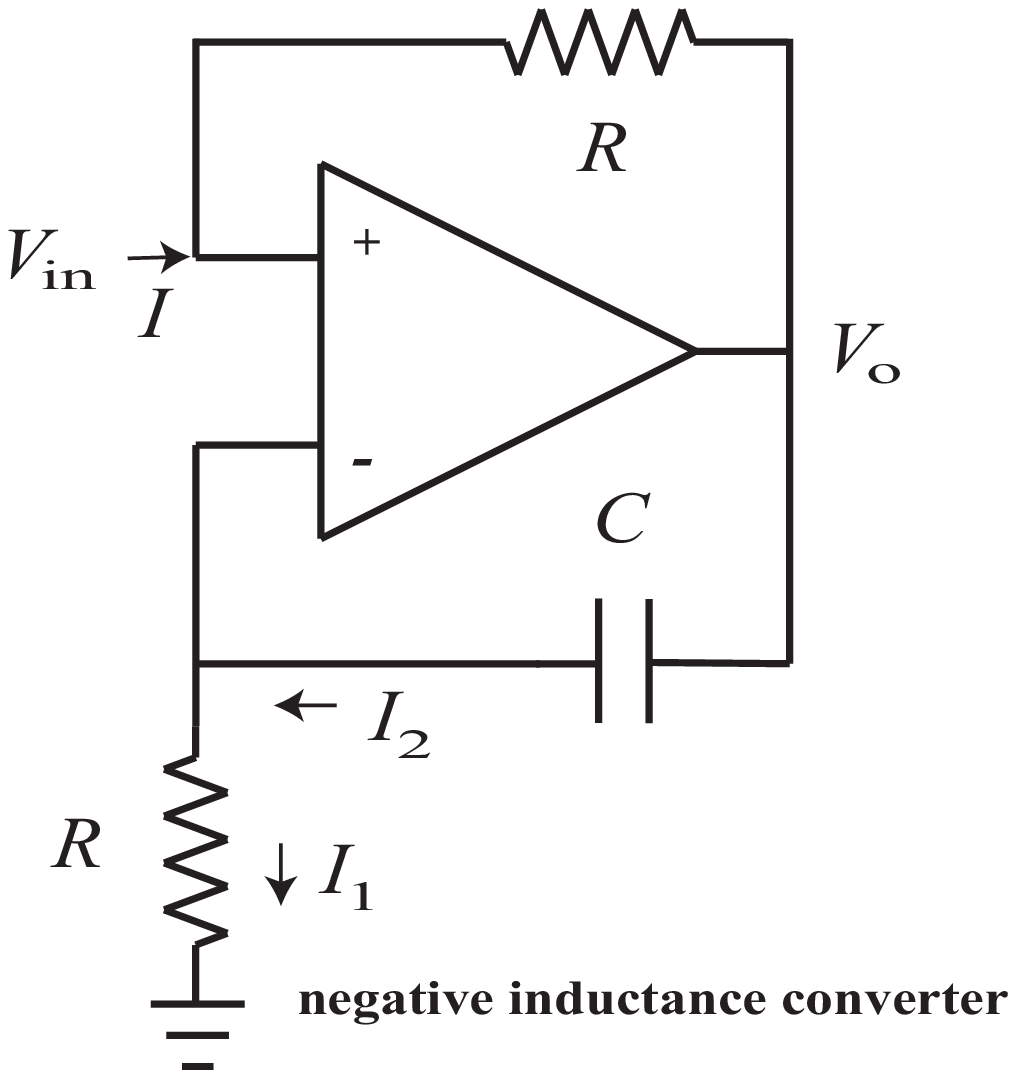}
\par\end{centering}
\begin{centering}
(a)\hspace*{5cm}(b)\hspace*{5cm}(c)
\par\end{centering}
\centering{}\caption{\label{fig:neg-imp} Operational-amplifier-based negative (a) impedance
converter; (b) capacitance converter; (c) inductance converter.}
\end{figure}

There is a number of physical devices that can provide for negative
capacitances and inductances needed for our circuits \cite[29]{Dorf}.
Figure \ref{fig:neg-imp} shows operational amplifiers that implement
negative impedance, capacitance and inductance respectively, Following
to \cite[10]{Iza}.
The currents and voltages for circuits depicted in Fig. \ref{fig:neg-imp}
are respectively as follows: (i) for negative impedance as in Fig.
\ref{fig:neg-imp}(a)

\begin{equation}
V_{\mathrm{in}}=-ZI,\quad V_{\mathrm{o}}=2V_{\mathrm{in}},\quad I_{1}=I_{2}=\frac{V_{\mathrm{in}}}{R};\label{eq:opamp1a}
\end{equation}
(ii) for negative capacitance as in Fig. \ref{fig:neg-imp}(b)

\begin{gather}
V_{\mathrm{in}}=Z_{\mathrm{in}}I,\quad Z_{\mathrm{in}}=-\frac{\mathrm{i}}{\omega C},\quad V_{\mathrm{o}}=2V_{\mathrm{in}},\quad I_{1}=I_{2}=\frac{V_{\mathrm{in}}}{R};\label{eq:opamp1b}
\end{gather}
(iii) for negative inductance as in Fig. \ref{fig:neg-imp}(c) 

\begin{gather}
V_{\mathrm{in}}=Z_{\mathrm{in}}I,\quad Z_{\mathrm{in}}=-\mathrm{i}\omega R^{2}C,\quad I_{1}=I_{2}=\frac{V_{\mathrm{in}}}{R},\quad V_{\mathrm{o}}=V_{\mathrm{in}}\left(1+\frac{1}{\mathrm{i}\omega RC}\right).\label{eq:opamp1c}
\end{gather}

\subsection{Topological aspects of the electric networks\label{subsec:net-top}}

We follow here mostly to \cite[2]{BalBic}. The purpose of this section
is to concisely describe and illustrate relevant concepts with understanding
that the precise description of all aspects of the concepts is available
in \cite[2]{BalBic}.

To describe topological (geometric) features of the electric network
we use the concept of \emph{linear graph} defined as a collection
of points, called \emph{nodes}, and line segments called branches,
the nodes being joined together by the branches as indicated in Fig.
\ref{fig:cir-gyr} (b). Branches whose ends fall on a node are said
to be incident at the node. For instance, Fig. \ref{fig:cir-gyr}
(b) branches \textbf{\textit{1}}, \textbf{\textit{2}}, \textbf{\textit{3}},
\textbf{\textit{4}} are incident at node 2. Each branch in Fig. \ref{fig:cir-gyr}
(b) carries an arrow indicating its orientation. A graph with oriented
branches is called an oriented graph. The elements of a network associated
with its graph have both a voltage and a current variable, each with
its own reference. In order to relate the orientation of the branches
of the graph to these references the convention is made that the voltage
and current of an element have the standard reference - voltage-reference
``plus'' at the tail of the current-reference arrow. The branch
orientation of a graph is assumed to coincide with the associated
current reference as shown in Figures \ref{fig:cir-CLR} and \ref{fig:cir-gyr}.

We denote the number of branches of the network by $N_{\mathrm{b}}\geq2$,
and the number of nodes by $N_{\mathrm{n}}\geq2$. 

\begin{figure}[th]
\begin{centering}
\includegraphics[clip,scale=0.5]{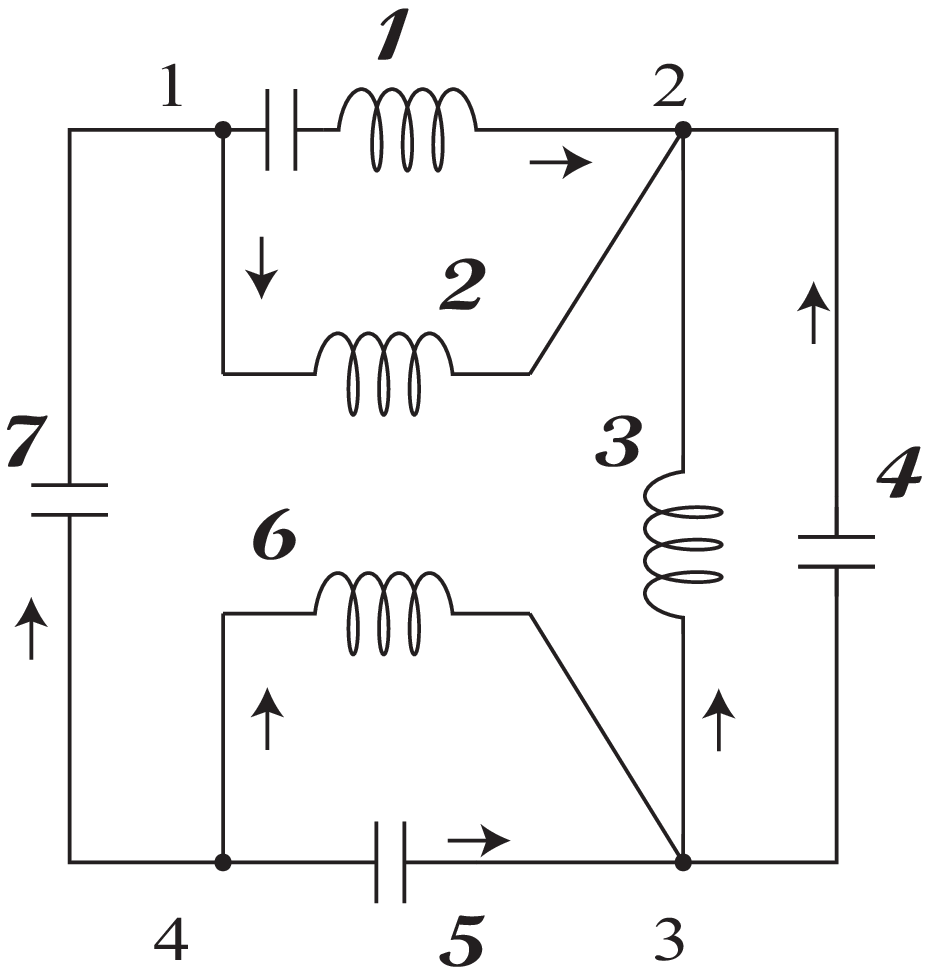}\hspace*{2cm}\includegraphics[clip,scale=0.5]{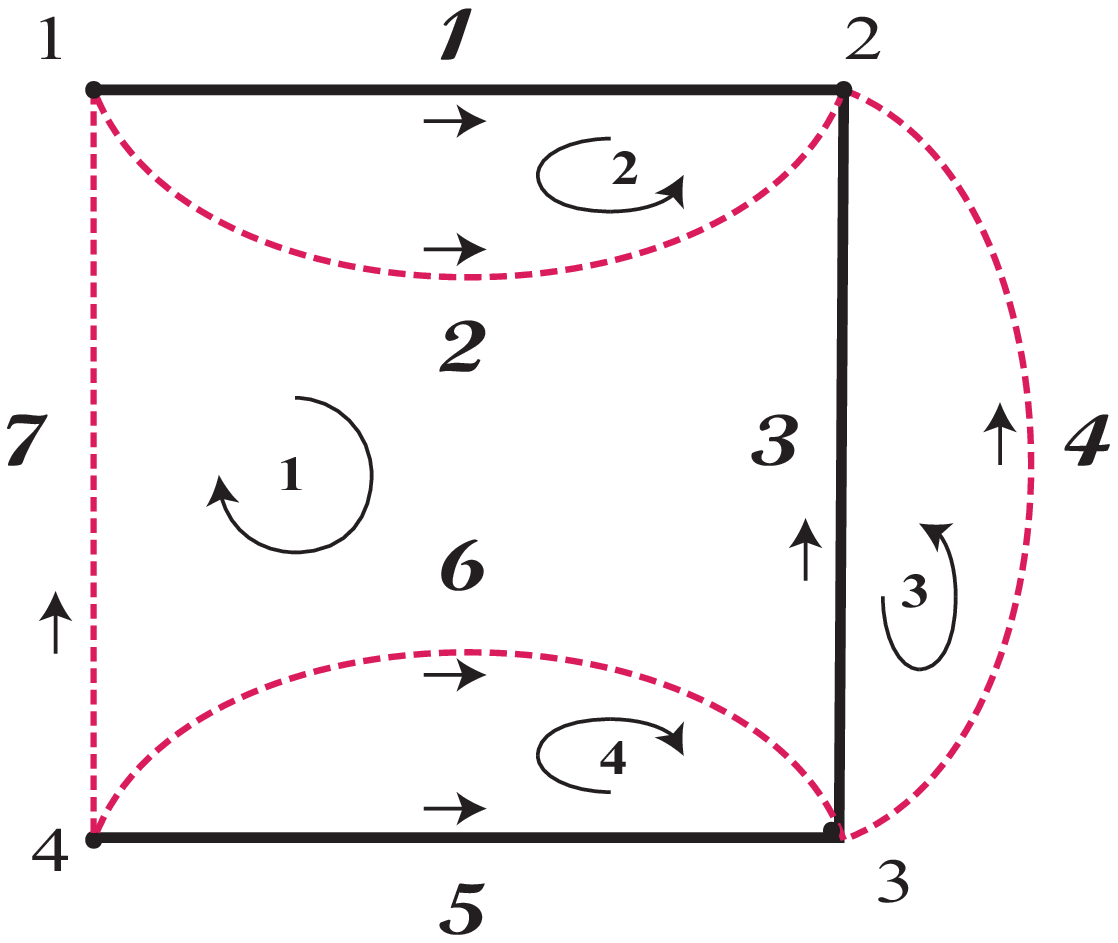}
\par\end{centering}
\centering{}\hspace*{-1cm}(a)\hspace*{6.5cm}(b)\caption{\label{fig:tree} The network (a) and its graph (b). There are $4$
nodes marked by small disks (black). In (b) there are $3$ twigs identified
by bolder (black) lines and labeled by numbers \textbf{\textit{1}},
\textbf{\textit{3}}, \textbf{\textit{5}}. There are $4$ links identified
by dashed (red) lines and labeled by numbers \textbf{\textit{2}},
\textbf{\textit{4}}, \textbf{\textit{6}}, \textbf{\textit{7}}. There
also $4$ oriented f-loops formed by the branches as shown.}
\end{figure}

A \emph{subgraph} is a subset of the branches and nodes of a graph.
The subgraph is said to be \emph{proper} if it consists of strictly
less than all the branches and nodes of the graph. A \emph{path} is
a particular subgraph consisting of an ordered sequence of branches
having the following properties:
\begin{enumerate}
\item At all but two of its nodes, called internal nodes, there are incident
exactly two branches of the subgraph. 
\item At each of the remaining two nodes, called the terminal nodes, there
is incident exactly one branch of the subgraph. 
\item No proper subgraph of this subgraph, having the same two terminal
nodes, has properties 1 and 2.
\end{enumerate}
A graph is called \emph{connected} if there exists at least one path
between any two nodes. We consider here only connected graphs such
as shown in Fig. \ref{fig:tree} (b).

A \emph{loop} (cycle) is a particular connected subgraph of a graph
such that at each of its nodes there are exactly two incident branches
of the subgraph. Consequently, if the two terminal nodes of a path
coincide we get a ``closed path'', that is a loop. In Fig. \ref{fig:tree}
(b) branches \textbf{\textit{7}}, \textbf{\textit{1}}, \textbf{\textit{3}},
\textbf{\textit{5}} together with nodes 1, 2, 3, and 4 form a loop.
We can specify a loop by an either the ordered list of the relevant
branched or the ordered list of the relevant nodes.

We remind that each branch of the network graph is associated with
two functions of time $t$: its current $I(t)$ and its voltage $V(t)$.
The set of these functions satisfy two Kirchhoff's laws, \cite[2.2]{BalBic},
\cite[2]{Cau}, \cite[Circuit Theory]{Rich}, \cite[1]{SesRee}. The
\emph{Kirchhoff current law} (KCL) states that in any electric network
the sum of all currents leaving any node equals zero at any instant
of time. The \emph{Kirchhoff voltage law} (KVL) states that in any
electric network, the sum of voltages of all branches forming any
loop equals zero at any instant of time. It turns out that the number
of independent KCL equations is $N_{\mathrm{n}}-1$ and the number
KVL equations is $N_{\mathrm{fl}}=N_{\mathrm{b}}-N_{\mathrm{n}}+1$
(the first Betti number \cite[2]{Cau}, \cite[2.3]{SesRee}).

There is an important concept of a \emph{tree} in the network graph
theory \cite[2.2]{BalBic}, \cite[2.1]{Cau} and \cite[2.3]{SesRee}.
A \emph{tree}, known also as \emph{complete tree}, is defined as a
connected subgraph of a connected graph containing all the nodes of
the graph but containing no loops as illustrated in Fig. \ref{fig:tree}
(b). The branches of the tree are called \emph{twigs} and those branches
that are not on a tree are called \emph{links} \cite[2.2]{BalBic}.
The links constitute the complement of the tree, or the \emph{cotree}.
The decomposition of the graph into a tree and cotree is not a unique.

The \emph{system of fundamental loops} or \emph{system of }f-loops
for short, \cite[2.2]{BalBic}, \cite[2.1]{Cau} and \cite[2.3]{SesRee},
is of particular importance to our studies. The system of time-dependent
charges (defined as the time integrals of the currents) associated
with the system of f-loops provides a complete set of independent
variables. When the network tree is selected then every link defines
the containing it f-loop. The orientation of an f-loop is defined
by the orientation of the link it contains. Consequently, there are
as many of f-loops in as there are links, and
\begin{equation}
\text{number of \ensuremath{f}-loops}:\;N_{\mathrm{fl}}=N_{\mathrm{b}}-N_{\mathrm{n}}+1.\label{eq:numfL}
\end{equation}
The number $N_{\mathrm{fl}}$ of f-loops defined by equation (\ref{eq:numfL})
quantifies the connectivity of the network graph, and it is known
in the algebraic topology as the first Betti number \cite[2]{Cau},
\cite[2.3]{SesRee}), \cite{Witt}.

The discussed concepts of the graph of an electric network such as
the tree, twigs, links and f-loops are illustrated in Fig. \ref{fig:tree}.
In particular, there are $4$ nodes marked by small disks (black).
In Fig. \ref{fig:tree} (b) there are $3$ twigs identified by bolder
(black) lines and labeled by numbers \textbf{\textit{1}}, \textbf{\textit{3}},
\textbf{\textit{5}}. There are $4$ links identified by dashed (red)
lines and labeled by numbers \textbf{\textit{2}}, \textbf{\textit{4}},
\textbf{\textit{6}}, \textbf{\textit{7}}. There also $4$ oriented
f-loops formed by the branches as follows: (1) \textbf{\textit{7}},
\textbf{\textit{1}}, \textbf{\textit{3}}, \textbf{\textit{5}}; (2)\textbf{\textit{
2}}, \textbf{\textit{1}}; (3)\textbf{\textit{ 4}}, \textbf{\textit{3}};
(2)\textbf{\textit{ 6}}, \textbf{\textit{5}}. These representations
of the f-loops as ordered lists of branches identify the corresponding
links as number in the first position in every list.

One also distinguishes simpler \emph{planar} networks with graphs
that can be drawn so that lines representing branches do not intersect.
The graph of a general electric network does not have to be planar
though. Networks with non-planar graphs can still be represented graphically
with more complex display arrangements or algebraically by the \emph{incidence
matrices}, \cite[2.2]{BalBic}.

\section{Conclusions}

We developed here an algorithm of a synthesis of a lossless electric
circuit based on prescribed Lagrangian. This means that the electric
circuit evolution equations are equivalent to the relevant Euler-Lagrange
equations. The synthesized circuit is composed of (i) capacitors and
inductors of positive or negative values for respective capacitances
and inductances, and (ii) gyrators. The proposed synthesis can be
viewed as a systematic approach in a realization of any finite dimensional
physical system described by a Lagrangian in a lossless electric circuit.

The synthesis can be used in a number of ways. It can be used to realize
the desired spectral properties in an electric circuit through a Lagrangian
that carries the properties directly. It can be also used in the realization
of arbitrary mutual capacitances and inductances in terms of elementary
capacitors and inductors of positive and negative respective capacitances
and inductances. The synthesis can be utilized also to generate circuit
approximations to transmission lines and waveguides within the frequency
limitations. 
\begin{quotation}
\textbf{\vspace{0.2cm}
}
\end{quotation}
\textbf{Acknowledgment:} This research was supported by AFOSR grant
\# FA9550-19-1-0103 and Northrop Grumman grant \# 2326345.

We are grateful to Prof. F. Capolino, University of California at
Irvine, for reading the manuscript and giving valuable suggestions.

Data sharing is not applicable to this article as no data sets were
generated or analyzed during the current study.

\section{Appendix: Concise review of the Lagrangian formalism\label{sec:Lag}}

Following common practice in the circuit theory, we assume the circuit
system configuration to be described by $N$-dimensional vector-columns
of voltages $V(z,t)=\left\{ V_{s}(z,t)\right\} _{s=1,\ldots,N}$ and
currents $I(z,t)=\left\{ I_{s}(z,t)\right\} _{s=1,\ldots,N}$. Our
primary circuit variables though are $N$-dimensional vector-column
of charges
\begin{gather}
Q(z,t)=\left\{ Q_{s}(z,t)\right\} _{s=1,\ldots,N},\quad Q_{s}(z,t)=\int^{t}I_{s}(z,t^{\prime})\,\mathrm{d}t^{\prime}.\label{eq:QsIn}
\end{gather}

The Lagrangian $\mathcal{L}$ for a linear system is a quadratic function
(bilinear form) of the system state $Q=\left[q_{r}\right]_{r=1}^{n}$
(column vector) and its time derivatives $\partial_{t}Q$, that is
\begin{gather}
\mathcal{L}=\mathcal{L}\left(Q,\partial_{t}Q\right)=\frac{1}{2}\left[\begin{array}{l}
Q\\
\partial_{t}Q
\end{array}\right]^{\mathrm{T}}M_{\mathrm{L}}\left[\begin{array}{l}
Q\\
\partial_{t}Q
\end{array}\right],\quad M_{\mathrm{L}}=\left[\begin{array}{rr}
-\eta & \theta^{\mathrm{T}}\\
\theta & \alpha
\end{array}\right],\label{eq:dlag1}
\end{gather}
where $\mathrm{T}$ denotes the matrix transposition operation, and
$\alpha,\eta$ and $\theta$ are $n\times n$-matrices with real-valued
entries. In addition to that, we assume matrices $\alpha,\eta$ to
be symmetric, that is 
\begin{equation}
\alpha=\alpha^{\mathrm{T}},\qquad\eta=\eta^{\mathrm{T}}.\label{eq:dlag2}
\end{equation}
Consequently,
\begin{equation}
\mathcal{L}=\frac{1}{2}\partial_{t}Q^{\mathrm{T}}\alpha\partial_{t}Q+\partial_{t}Q^{\mathrm{T}}\theta Q-\frac{1}{2}Q^{\mathrm{T}}\eta Q.\label{eq:dlag3a}
\end{equation}
Then by Hamilton's principle, the system evolution is governed by
the EL equations 
\begin{equation}
\frac{d}{dt}\left(\frac{\partial\mathcal{L}}{\partial\partial_{t}Q}\right)-\frac{\partial\mathcal{L}}{\partial Q}=0,\label{eq:dlag4}
\end{equation}
which, in view of equation (\ref{eq:dlag3a}) for the Lagrangian $\mathcal{L}$,
turns into the following second-order vector ordinary differential
equation (ODE):
\begin{equation}
\alpha\partial_{t}^{2}Q+\left(\theta-\theta^{\mathrm{T}}\right)\partial_{t}Q+\eta Q=0.\label{eq:dlag5}
\end{equation}
Notice that matrix $\theta$ enters equation (\ref{eq:dlag5}) through
its skew-symmetric component $\frac{1}{2}\left(\theta-\theta^{\mathrm{T}}\right)$
justifying as a possibility to impose the skew-symmetry assumption
on $\theta$, that is
\begin{equation}
\theta^{\mathrm{T}}=-\theta.\label{eq:dlag3aa}
\end{equation}
Indeed, the symmetric part $\theta_{s}=\frac{1}{2}\left(\theta+\theta^{\mathrm{T}}\right)$
of the matrix $\theta$ is associated with a term to the Lagrangian
which can be recast as is the complete (total) derivative, namely
$\frac{1}{2}\partial_{t}\left(Q^{\mathrm{T}}\theta_{s}Q\right)$.
It is a well known fact that adding to a Lagrangian the complete (total)
derivative of a function of $Q$ does not alter the EL equations.
Namely, the EL equations are invariant under the Lagrangian gauge
transform $\mathcal{L}\rightarrow\mathcal{L}+\partial_{t}F\left(q,t\right)$,
\cite[2.9, 2.10]{Scheck}, \cite[I.2]{LanLifM}.

Under the assumption (\ref{eq:dlag3aa}) equation (\ref{eq:dlag5})
turns into its version with the skew-symmetric $\theta$
\begin{equation}
\alpha\partial_{t}^{2}Q+2\theta\partial_{t}Q+\eta Q=0,\text{ if }\theta^{\mathrm{T}}=-\theta.\label{eq:dlag5aa}
\end{equation}
It turns out though that the Lagrangian that corresponds to the Hamiltonian
by the Legendre transformation does not have to have skew-symmetric
$\theta$ satisfying (\ref{eq:dlag3aa}). For this reason we don't
impose the condition of skew-symmetry on $\theta$. 

\subsection{Lagrangian framework for an electric network\label{subsec:cir-Lag}}

The possibility of the construction of the Lagrangian framework for
electric circuits is well known. Basic examples of such a construction
are provided, for instance, in \cite[9]{GantM}, \cite[15]{Wells},
The well-known expressions for the Lagrangian for the basis circuit
elements considered in Section \ref{subsec:cir-elem} are as follows
\cite[9]{GantM}, \cite[3]{Rich}:
\begin{align}
\text{capacitance: }\mathcal{L} & =-\frac{\left(Q\right)^{2}}{2C},\label{eq:LagCL1a}
\end{align}
\begin{align}
\text{inductance: }\mathcal{L} & =\frac{L\left(\partial_{t}Q\right)^{2}}{2},\label{eq:LagCL1b}
\end{align}
\begin{align}
\text{gyrator: }\mathcal{L} & =\frac{G}{2}\left(Q_{1}\partial_{t}Q_{2}-Q_{2}\partial_{t}Q_{1}\right),\label{eq:LagCL1c}
\end{align}

Suppose now that we have a lossless circuit composed of capacitors,
inductors and gyrators of positive or negative values for the respective
capacitances and inductances. There is a well defined procedure (algorithm)
that allows to assign to such a circuit a Lagrangian. This procedure
is as follows.

The first immediate question as we start the construction of the circuit
Lagrangian is what is the set of independent variables describing
the state of the circuit at any point of time? To answer this question
we use the results of Section \ref{subsec:net-top} and identify a
circuit tree and the corresponding set of f-loops. With that been
done we introduce the charges $Q_{m}$ and the currents $\partial_{t}Q_{m}$,
$1\leq m\leq N$ associated the corresponding $N$ f-loops to be the
generalized coordinates defining the state of the circuit at any point
of time.

Then the process of generating the circuit Lagrangian $\mathcal{L}$
follows to the following steps.
\begin{enumerate}
\item If $\mathrm{b}$ is a link, see the definition in Section \ref{subsec:net-top},
associated with the f-loop of the index $m=m\left(\mathrm{b}\right)$
then it is associated with either capacitance $C_{\mathrm{b}}$ or
inductance $L_{\mathrm{b}}$ and is assigned respectively according
to equations (\ref{eq:LagCL1a}) and (\ref{eq:LagCL1b}) the following
value of the Lagrangian
\begin{equation}
\mathcal{L}_{\mathrm{b}}=\frac{Q_{m}^{2}}{2C_{\mathrm{b}}},\quad\mathcal{L}_{\mathrm{b}}=\frac{L_{\mathrm{b}}\left[\partial_{t}Q_{m}\right]^{2}}{2},\;m=m\left(\mathrm{b}\right).\label{eq:LagCL2a}
\end{equation}
\item If $\mathrm{b}$ is a twig, see the definition in Section \ref{subsec:net-top},
we consider the set $S\left(\mathrm{b}\right)$ of indexes $m$ of
f-loops containing the twig $\mathrm{b}$. To account for mutual orientations
of branches and f-loops and consequently the signs of f-loop currents
we introduce for any $m\in S\left(\mathrm{b}\right)$ the corresponding
$\epsilon\left(m\right)=\pm1$ as follows. If the orientations of
the twig $\mathrm{b}$ and the $m$-th f-loop that includes it are
the same we set $\epsilon\left(m\right)=1$, otherwise $\epsilon\left(m\right)=-1$.
If the twig is associated with either capacitance $C_{\mathrm{b}}$
or inductance $L_{\mathrm{b}}$ we assign to it respectively the following
Lagrangian
\begin{gather}
\mathcal{L}_{\mathrm{b}}=\frac{Q_{\mathrm{b}}^{2}}{2C_{\mathrm{b}}},\quad\mathcal{L}_{\mathrm{b}}=\frac{L_{\mathrm{b}}\left[\partial_{t}Q_{\mathrm{b}}\right]^{2}}{2},\quad Q_{\mathrm{b}}=\sum_{m\in S\left(\mathrm{b}\right)}\epsilon\left(m\right)Q_{m}.\label{eq:LagCL2b}
\end{gather}
\item As to the gyrators notice that every gyrator by its very design couples
a pair of some two loops. With that in mind we consider all pairs
of f-loops with indexes $1\leq m<k\leq N$ and the corresponding gyration
resistances $G_{km}$ defined by equations (\ref{eq:LagGLC1ea}).
If a pair of f-loops is not coupled the corresponding gyration resistance
is set to be zero. Then for each pair $m$ and $k$ we use equation
(\ref{eq:LagCL1c}) and define the corresponding Lagrangians $\mathcal{L}_{km}$
by the following equations:
\begin{gather}
\mathcal{L}_{km}=\frac{G_{km}}{2}\left(Q_{m}\partial_{t}Q_{k}-Q_{k}\partial_{t}Q_{m}\right),\quad1\leq m<k\leq N.\label{eq:LagCL2c}
\end{gather}
\item Using the Lagrangian components $\mathcal{L}_{\mathrm{b}}$ and $\mathcal{L}_{km}$
defined in previous steps we the define circuit Lagrangian as their
sum, that is
\begin{equation}
\mathcal{L}=\sum_{\mathrm{b}\in\mathrm{B}}\mathcal{L}_{\mathrm{b}}+\sum_{1\leq m<k\leq N}\mathcal{L}_{km},\label{eq:LagCL2d}
\end{equation}
where $\mathrm{B}$ is the set of all branches, that is all the links
and the twigs.
\end{enumerate}
A straightforward examination of the Euler-Lagrange equations for
the circuit Lagrangian $\mathcal{L}$ defined by equations (\ref{eq:LagCL2d})
confirms the well-known fact that they can be interpreted as Kirchhoff's
voltage law, that is the sum of all voltages around a loop (closed
circuit) is zero.

\end{document}